\documentclass[useAMS,usenatbib]{mn2e}
\usepackage{amsmath}
\usepackage{graphicx}
\usepackage{subfigure}

\begin{document}

\title{Unbound Star-forming Molecular Clouds}

\author[Ward, Wadsley, \& Sills]{Rachel L. Ward, James Wadsley, and Alison Sills \\
 Department of Physics and Astronomy, McMaster University, Hamilton, ON, L8S 4M1, Canada}

\label{firstpage}

\maketitle

\begin{abstract} 
We explore whether observed molecular clouds could include a substantial population of unbound clouds.  Using simulations which include only turbulence and gravity, we are able to match observed relations and naturally reproduce the observed scatter in the cloud size-linewidth coefficient, at fixed surface density.  We identify the source of this scatter as a spread in the intrinsic virial parameter.  Thus these observational trends do not require that clouds exist in a state of dynamical equilibrium.  We demonstrate that cloud virial parameters can be accurately determined observationally with an appropriate size estimator.  All our simulated clouds eventually form collapsing cores, regardless of whether the cloud is bound overall.   This supports the idea that molecular clouds do not have to be bound to form stars or to have observed properties like those of nearby low-mass clouds.
\end{abstract}


\section{Introduction}\label{sec:Intro}

Molecular clouds are the principal sites of star formation, as localised dense regions of gas and dust within a cold turbulent cloud collapse under gravity to form stars and stellar clusters.  These highly supersonic clouds are typically tens of parsecs across, reaching masses of up to 10$^6$ M$_{\odot}$ in the Milky Way.  In order to reach a greater understanding of how stars form and evolve, a logical first step is to explore the large-scale structure and dynamics of molecular clouds.  

One approach is to characterise molecular clouds based on their large-scale properties: mass, size, and velocity dispersion.  
For the past thirty years, the three scaling relations presented in \citet{larson81}, commonly called Larson's Laws, have been used to link these properties.  Larson found an empirical relation between the velocity dispersion and size of molecular clouds such that $\sigma_{3\text{D}}$ = 1.1 L$^{0.38}$ where $\sigma_{3\text{D}}$ is the three-dimensional internal velocity dispersion in km s$^{-1}$ and L is the maximum linear dimension in pc.  \citet{solomon87} used a large sample of $^{12}$CO observations of molecular clouds to find a similar size-linewidth relation of $\sigma_{v}$ = (1.0 $\pm$ 0.1) S$^{0.5}$ where $\sigma_{v}$ is the one-dimensional velocity dispersion in km s$^{-1}$ and S is a measure of the cloud size in pc.  Although their results are consistent with Larson's, they do find a steeper power-law index of 0.5 which the authors argue is a consequence of clouds in virial equilibrium rather than due to a Kolmogorov turbulent spectrum, as suggested by \citet{larson81}.  This interpretation by \citet{solomon87} is also consistent with Larson's second conclusion that molecular clouds are gravitationally bound and in approximate virial equilibrium.  The virial state of a cloud is often expressed using the virial parameter, $\alpha$, 
\begin{equation}
   \alpha = \frac{5\sigma_{v}^2\text{R}}{\text{GM}}, 
\label{eq:alphadef}
\end{equation}
\citep[e.g.][]{BM92} where $\sigma_{v}$ is the one-dimensional velocity dispersion along the line of sight, R is the maximum projected radius, and M is the total mass of the cloud.  The virial parameter describes simple virial equilibrium which excludes effects of magnetic fields and external pressure.  Spherical, homogeneous clouds are virialised when $\alpha$ = 1 and gravitationally bound if $\alpha$ $\leq$ 2.
Larson's third scaling relation proposed that molecular clouds have approximately the same column density on all size scales; however, this conclusion was based on a sample with a limited range of column densities, which was the best achievable by observations at the time.  If molecular clouds are assumed to be in gravitational equilibrium with a constant column density, then the size-linewidth relation naturally follows, implying the universality of molecular cloud structure and turbulence.

Although all three of Larson's scaling relationships are frequently used to define the characteristics of molecular clouds, there are often large amounts of scatter accompanying each trend, which varies in many cases by over two orders of magnitude \citep[e.g.][]{larson81, balle2011a,wong2011,hcs01}.  The scaling relations are often the focus of study while the scatter itself is readily dismissed.  Whether this scatter is physically meaningful or a product of observational uncertainty is a subject which is explored throughout this work and also most notably in an influential study by \citet{heyer2009} involving a re-examination of the \citet{solomon87} cloud sample using $^{13}$CO observations with greater sensitivity and higher spectral and angular resolution.  Their results showed that the cloud column density is not constant, as confirmed in later studies \citep[e.g.][]{lombardi2010, kauf2010b, balle2011a}, and that the scatter present in their size-linewidth relation is reduced when the scaling coefficient depends on the cloud mass surface density, $\Sigma$.  The size-linewidth scaling coefficient, defined by \citet{heyer2009} as
\begin{equation}
   v_{\text{o}} = \sigma_v/\text{R}^{1/2} = (\pi \text{G} \Sigma/5)^{1/2},
\label{eq:coefficient}
\end{equation}
indicates that clouds inhabit a one-dimensional space parameterized by surface density.  \citet{heyer2009} suggested that the dependence of the size-linewidth coefficient, $v_{\text{o}}$, on the surface density may not have been realized previously due to the limited range of column densities which could be probed in early observations of molecular clouds.  The authors also note that scatter, while reduced, is still present in their data for the size-linewidth coefficient vs. the surface density. 

A number of possible explanations for the cause of this remaining scatter have been proposed.  \citet{heyer2009} suggested that different interstellar magnetic field strengths and varying flux-to-mass ratios could produce the observed trend and scatter since the relation is consistent with models of magnetically-supported molecular clouds \citep{mous87}.  However, since the magnetic field strengths are unknown for this sample of molecular clouds, this cannot be confirmed by current observations.  Alternatively, \citet{field11} found that the data of \citet{heyer2009} can be accounted for by using different external pressure values for clouds in pressure-bounded virial equilibrium; however, the range of values necessary to reproduce the observed trend extends much higher than current estimates for pressures present in the neutral ISM \citep{elmegreen89,BM92}. 

\citet{balle2011a} argue that neither magnetic support nor pressure confinement are required to explain the relation and scatter as the data are simply consistent with clouds undergoing hierarchical and chaotic gravitational collapse.  Molecular clouds would have localised collapsing regions within a globally bound turbulent cloud, but these regions may not necessarily be virialised at all stages throughout their evolution.  Although \citet{heyer2009} find that most of the clouds in their sample are super-virial, they remark that their derived cloud masses could have been underestimated by a factor of 2 -- 3.  However, \citet{dobbs11} showed that even after doubling the cloud masses, most of the clouds in the \citet{heyer2009} sample are not virialised with 50\% of the clouds strictly unbound with $\alpha$ $>$ 2 in contrast to the common assumption of bound clouds.

The dependence of the size-linewidth coefficient on the mass surface density is expected from the definition of the virial parameter for clouds assumed to be gravitationally bound and virialised ($\alpha$ = 1).  However, there is considerable scatter present and the most direct interpretation of this scatter implies unbound clouds, $\alpha > 2$.  Molecular clouds which are unbound as a whole can contain dense gravitationally bound subregions within them, such as clumps and cores, initially formed by supersonic turbulent motions and maintained by their own self-gravity.  Whether or not molecular clouds can be truly unbound and still consistent with observations is the primary focus of this work.    

Simulators have begun to explore the possibility that molecular clouds are unbound.  Recent simulations \citep{clark05,clark08,bonnell2011} have shown that unbound clouds can result in localised and distributed star formation and produce naturally low star formation efficiencies.  \citet{dobbs11} explored the formation of molecular clouds through galaxy-scale simulations.  Their simulated clouds were able to remain unbound, which the authors attributed to cloud-cloud collisions and stellar feedback.  They produce molecular clouds with a wide range of virial parameters; however, they were not able to follow internal structure or dynamics in the clouds.  Furthermore, their resolution prevents them from probing more typical cloud masses ($\la$ 10$^{5}$ M$_{\odot}$) such as those in the \citet{heyer2009} sample.   

Our simulations have been designed to avoid potential biases associated with assuming molecular clouds are initially bound, that they have a preferred surface density, or other assumptions that predispose them to collapse and star formation.  We produced a set of 16 high resolution simulations, with properties listed in Table~\ref{table:initcondit}, that cover a broad range of initial column densities and turbulent velocities.  Our simulations evolve to form local regions with surface densities typical of observed molecular clouds.  Unlike most other studies, the clouds also contain large-scale turbulent motions so they do not collapse spherically.  In this paper, we present properties of these clouds as they would be observed.   To create a sample of synthetic observations at various stages of evolution, we selected localised dense turbulent regions analogous to low-mass clouds, such as the Taurus molecular cloud \citep[1.5 $\times$ 10$^4$ M$_{\odot}$;][]{pineda10} and the Perseus molecular cloud \citep[7 $\times$ 10$^3$ M$_{\odot}$;][]{arce2010}.  We demonstrate the ability of these clouds to form stars, and we offer a simple explanation for the scatter present in the $v_{\text{o}}$-$\Sigma$ relation.  

In Section~\ref{sec:Sim}, we describe the details of our suite of simulations followed by a description of our cloud selection criteria and methods of analysis in Section~\ref{sec:sample}.  We present our results and explore the effect of boundedness on the determination of cloud properties in Section~\ref{sec:results}.  Finally, we summarize our results in Section~\ref{sec:discussion}.

\section{The Simulations}\label{sec:Sim}

Using the smoothed particle hydrodynamics code, \textsc{Gasoline} \citep{gasoline}, we ran sixteen simulations of a 50 000 M$_{\odot}$ region of the ISM which include the effects of decaying initial turbulence and gravity.  This allows us to study the dynamics and evolution of molecular clouds without the additional complications of magnetic fields, variable external pressure, or stellar feedback.  
  
The simulations use the same equation of state and open boundary conditions as those described in \citet{ward12a}.  Since our equation of state from \citet{bb05} has an opacity limit for fragmentation of 10$^{-13}$ g cm$^{-3}$ much higher than our maximum gas density, our entire simulations are optically thin and isothermal at 10 K.  Each run is simulated using approximately 5.5 million particles, resulting in one of the most massive and highly resolved simulations of an isolated cloud currently achieved, with a mass resolution of 8.85 $\times$ 10$^{-3}$ M$_{\odot}$ per particle.  

We used a Burgers' turbulent velocity spectrum characterised by three eigenvectors of the symmetric shear tensor, where material along one axis expanded outward, material along another axis collapsed inward, and the third dimension remained static.  We refer to this type of anisotropic collapse as the `ribbon' collapse case \citep{nic, wardthesis}, which was determined by \citet{nic} through a statistical study of the anisotropy of cloud collapse to be the most likely scenario to be observed in nature, assuming that all molecular clouds have random turbulence on large-scales.  Our simulations include turbulent modes larger than the scale of the cloud to account for large-scale effects.  The resultant flow structure has some similarities with that of a cloud formed from a colliding flow.  It also has a very filamentary structure which is in agreement with current observations of molecular clouds \citep[e.g.][]{sasha,arzou}.  In order to avoid the high computational costs of simulating such a large range of densities typical of observed clouds, the highest density regions ($>$ 10$^6$ atoms cm$^{-3}$) were replaced by sink particles with the requirement that these regions were at a potential minimum and were gravitationally bound \citep[cf.][]{federrath10}.  The simulations are run until a maximum of 35\% in mass of the gas is converted to stars.  Since several of our runs are initially unbound, the mass fraction in stars is not directly coupled to the free-fall time, resulting in different absolute stopping times for each run.  However, each simulation was run until t = 0.5 t$_{ff}$ at the very least, where t$_{ff}$ = 1/$\sqrt{G\rho_{\text{initial}}}$.
  
The initial conditions of our simulations were chosen such as to have initial physical virial parameters ranging from $\alpha_{\text{initial}}$ = 1 (bound and virialised) to $\alpha_{\text{initial}}$ = 10 (highly unbound).  This was achieved using a range of densities, n$_{\text{initial}}$, and velocity dispersions, $\sigma_{\text{3D}}$, listed in Table~\ref{table:initcondit}, resulting in column densities between 1 -- 7 $\times$ 10$^{21}$ cm$^{-2}$.  For comparison, observed molecular clouds have a mean extinction of A$_{\text{v}}$ $\approx$ 1 which corresponds to a column density of $\sim$ 10$^{21}$ cm$^{-2}$ \citep{bohlin}.  Although infrared dark clouds can reach column densities of $\sim$ 10$^{24}$ cm$^{-2}$ \citep{balle2011a}, near-infrared dust extinction data for molecular clouds can only probe to a maximum column density of $\sim$ 25 $\times$ 10$^{21}$ cm$^{-2}$ due to the observational limitations of this method (\citet{kain2009}, and references therein).  

Our simulations of the 50 000 M$_{\odot}$ region of the ISM evolve under gravity and turbulence to form molecular clouds with masses ranging from $\sim$ 300 to 10 000 M$_{\odot}$.  Detailed properties of these clouds and the process by which they were selected are provided in Section~\ref{sec:sample}.

\begin{table}
	\caption{Initial Conditions for the Simulations}
	\centering
	\begin{tabular}{c c c c c c c c}
	\\
	Id &  Radius    & $\sigma_{3\text{D}}$ & n$_{\text{initial}}$ & t$_{ff}$ & $\alpha_{\text{initial}}$\\
		         & (pc) & (km s$^{-1}$) & (cm$^{-3}$) & (Myr) &  \\
	\hline\hline   
	1  & 30 & 6.56 & 7.69 & 22.4 & 10\\
	2  & 21.2 & 7.80 & 21.8 & 13.3 & 10\\
	3  & 15 & 9.27 & 61.5 & 7.9 & 10\\
	4  & 10.6 & 11.0 & 174 & 4.7 & 10\\
	5  & 30 & 4.39 & 7.69 & 22.4 & 4\\ 
	6  & 21.2 & 5.22 & 21.8 & 13.3 & 4 \\
	7  & 15 & 6.20 & 61.5 & 7.9 & 4\\
	8  & 10.6 & 7.38 & 174 & 4.7 & 4\\
	9  & 30 & 2.93 & 7.69 & 22.4 & 2\\
	10 & 21.2 & 3.49 & 21.8 & 13.3 & 2\\
	11 & 15 & 4.15 & 61.5 & 7.9 & 2\\
	12 & 10.6 & 4.93 & 174 & 4.7 & 2\\
  13  & 30 & 2.07 & 7.69 & 22.4 & 1\\
  14 & 21.2 & 2.47 & 21.8 & 13.3 & 1\\
  15 & 15 & 2.93 & 61.5 & 7.9 & 1\\
  16 & 10.6 & 3.49 & 174 & 4.7 & 1\\
	\hline
  \end{tabular}
  \label{table:initcondit}
\end{table}

\section{The Cloud Selection Process}\label{sec:sample}

\begin{figure} 
\begin{center}
\subfigure[Sample Map]{\label{fg:synobscomplex}\includegraphics[scale=0.58, angle=0, clip=true, trim=3.8cm 1.5cm 3.8cm 9mm]{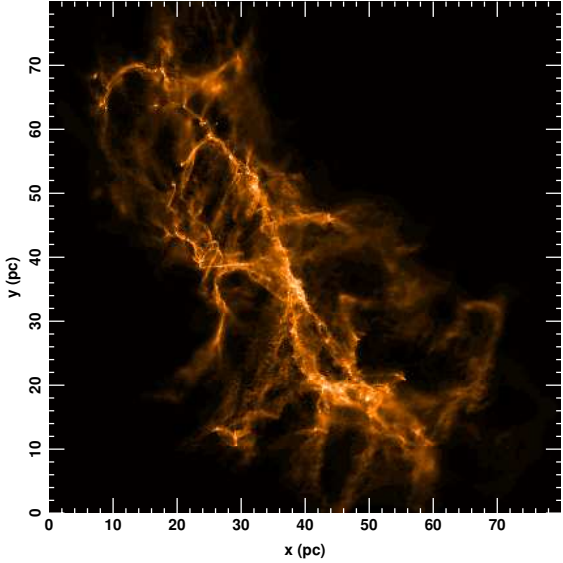}}
\subfigure[Smoothed Contour Map]{\label{fg:contourmap}\includegraphics[scale=0.48, angle=0, clip=true, trim=5cm 2.3cm 3.8cm 9mm]{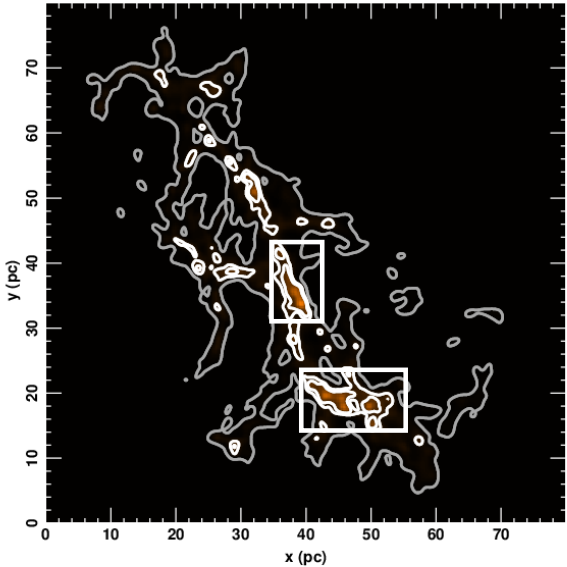}}
\caption{\label{fg:synobs}
  Sample synthetic column density maps for a marginally bound ($\alpha_{\text{init}}$ = 2) case with an initial radius of approximately 21 pc at 0.6 t$_{ff}$.  The top figure shows the full range of column density we can cover in our simulation.  The map has logarithmic scaling with a maximum column density of 25 $\times$ 10$^{21}$ cm$^{-2}$.  The figure on the bottom shows a map with a limited range of column density more consistent with observations.  The minimum contour level for consideration to be selected as a cloud (white contour) is chosen to be three times the value of our chosen threshold, $\sigma_{\text{th}}$ = 0.5 $\times$ 10$^{21}$ cm$^{-2}$ (grey contour).  The boxes identify the two clouds in this map which meet the selection criteria outlined in the text.
}
\end{center}
\end{figure}

In order to compare the results of the simulations with observations, we have created a synthetic two-dimensional column density map for each of the sixteen cases using the method described in \citet{ward12a}.  Our column density maps were produced from the simulation by interpolating the SPH particles to a 4000 x 4000 grid, where the spatial resolution is 6000 AU per pixel.  The SPH kernel function \citep{ML85} is used to smooth each particle over several grid cells and is weighted to ensure normalization of the total mass in the map.  A sample synthetic column density map is shown in Figure \ref{fg:synobscomplex}.  We can assume that the column density of material in our maps is a direct analogue of extinction measurements or of the observed flux from $^{13}$CO observations since our gas is optically thin for the duration of each simulation.  The cloud sample was created by examining each simulation at 1 Myr intervals for a total of 122 snapshots.  Individual clouds typically appear at around 1/4 of a free-fall time and sinks, representing a star or small groups of stars, begin to form at a similar time.  The sample therefore represents molecular clouds at many different stages of evolution.

Figure~\ref{fg:contourmap} demonstrates the method of cloud selection.  We temporarily smoothed our maps to 1 pc spatial resolution using a Gaussian filter to emphasise the large-scale structure.  This spatial resolution was chosen to be similar to the best resolution of the nearby molecular cloud observations by \citet{solomon87} and our selection process is intended to mimic theirs.  The grey contour lines mark the minimum possible column density, $\sigma_{\text{th}}$ $\approx$ 0.5 $\times$ 10$^{21}$ cm$^{-2}$, that can be detected from dust extinction measurements \citep{kain2009}.    
To be considered a cloud, the extended emission above the white contour lines, which have a minimum threshold of 3$\sigma_{\text{th}}$ and contour intervals of 3$\sigma_{\text{th}}$, must be contiguous with box dimensions greater than 3 pc across, where the size threshold distinguishes clouds from smaller star-forming clumps \citep{BT07}.  These newly-identified clouds were extracted from the simulations using a box size defined by the contour threshold in order to isolate and study local regions within the simulations.  This selection allows for a more direct comparison with clouds like the Taurus molecular cloud and the Pipe Nebula.  Our entire sample includes 181 clouds with masses ranging between 100 -- 10$^4$ M$_{\odot}$, consistent with observed masses of nearby clouds \citep{kfh13}.  Our population of clouds also spans a range of dynamical states, where some clouds are actively star-forming (localised collapse to densities greater than 10$^6$ cm$^{-3}$) and others are much more quiescent.  

\section{Results}\label{sec:results}

\subsection{Comparison to Observations}\label{subsec:cloudprops}

To ensure that the simulated molecular clouds are consistent with real molecular clouds, we compared properties determined from our synthetic column density maps to those measured in observations.  For the clouds in our sample, we only included pixels above a minimum column density threshold of 0.5$\times$10$^{21}$ cm$^{-2}$ in our determination of the mass, size, and linewidth.  This column density threshold corresponds to the lower limit for detection in extinction maps \citep[A$_{\text{v}}$ $\sim$ 0.5;][]{bohlin} which also coincides with the threshold for self-shielding against UV feedback leading to the formation of molecular clouds.

The cloud mass, M, is the total mass in the pixels above the threshold in each simulated cloud.  Our cloud masses are consistent with those measured in observations of nearby clouds \citep[median mass of nearby clouds = 5000 M$_{\odot}$;][]{kfh13}.  We have excluded sink particles from our estimations since the effects of stars are not included in observational estimates of molecular cloud properties. Embedded stars bias measurements in extinction-based data due to pixel `saturation' \citep[see][]{kain2009, goodman09}, and in molecular line observations, the use of CO and its isotopologues as tracers of H$_2$ cause `freeze-out' onto dust grains at densities of $\ga$ 10$^{4}$ cm$^{-3}$.  Also, \citet{kfh13} recently explored the effect of sink particles on the surrounding gas in their simulations of molecular clouds and argued in favour of their removal.  

We determined the line-of-sight velocity dispersions, $\sigma_{v_i}$, using the mass-weighted velocities of the gas particles in our simulation, $\langle v_i\rangle$, such that $\sigma_{v_i}^2 = \langle v_i^2 \rangle - \langle v_i \rangle^2$.  To avoid any dependence of the velocity dispersion on our single (arbitrarily chosen) line-of-sight, we used the averaged one-dimensional velocity dispersion to represent the linewidth:
\begin{equation}
   \sigma_{\text{1D}} = \left(\frac{\sum_i\sigma_{v_i}^2}{3}\right)^{1/2},\ \ \text{i = x, y, z}.
\end{equation}

The mass surface density is given by $\Sigma$ = M/$\pi$R$^2$, where R is the radius of the cloud; however, the radius can be quite difficult to determine as cloud boundaries can be subjective and dependent on the interpretation of the observer.  Molecular clouds are often first detected using CO observations \citep{wilson90, lada2009}, where the size is measured as the full extent of molecular emission \citep{solomon87}.  However, several authors argue that dust extinction maps are best for determining the spatial distribution of a cloud as they are affected by fewer uncertainties and can probe a larger dynamic range \citep{goodman09, kauf2010a, kain2009, lombardi2010, beaumont2012}. To determine which is a more appropriate estimate of size, we compare two means of measuring the cloud area using our sample of simulated clouds.

We initially defined the area of our clouds using the same method of \citet{solomon87} and \citet{heyer2009}, who used an intensity-weighted size dispersion to determine the angular extent of each molecular cloud.  As an analogy to intensity-weighting, we use the mass in each pixel to weight the size dispersions of our clouds, which are $\sigma_x$ and $\sigma_y$ in the xy-plane.  To relate the size parameter, S$_{\text{xy}} = \sqrt{\sigma_{x}\sigma_{y}}$, to the projected area of the cloud, S$_{\text{xy}}$ is scaled by a factor of 3.4 \citep{solomon87}, such that the effective radius, R$_{\text{S}}$, is defined to be 
\begin{equation}
		\text{R}_{\text{S}} = \frac{3.4}{\sqrt{\pi}} \text{S$_{\text{xy}}$}. \label{eq:solrad}
\end{equation}

\begin{figure}
\begin{center}
\includegraphics[scale=0.9, angle=0, clip=true, trim=0.5cm 0cm 0cm 0cm]{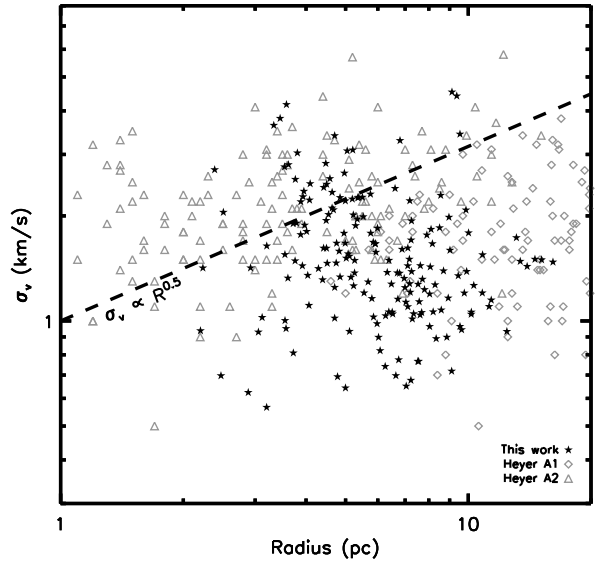}
\caption{\label{fg:sizelineplot}
Linewidth as a function of size, R$_{\text{S}}$, for simulated (black stars) and observed (grey symbols) molecular clouds.  The observational data was obtained from Table 1 of \citet{heyer2009} for clouds with areas determined by \citet{solomon87}, A$_1$ (open diamonds), and by the half-max isophote of peak column density, A$_2$ (open triangles). 
}
\end{center}
\end{figure}

Figure~\ref{fg:sizelineplot} shows the linewidth plotted as a function of radius as defined by \citet{solomon87}, R$_{S}$, for our clouds (black stars), as compared to real clouds (grey symbols) from the observations of \citet{heyer2009}.  \citet{heyer2009} use two methods of measuring area to determine their cloud properties: the primary area, A$_1$, defined by the boundaries used by \citet{solomon87} as described above and a secondary area, A$_2$, corresponding to the highest column density regions and defined by the area within the half-power isophote of the peak column density.  Both data sets are included in Figure~\ref{fg:sizelineplot} to demonstrate the range of size estimates made by observations.  Our clouds lie on the Larson size-linewidth relation (black dashed line) and are shown to have comparable velocity dispersions to observed clouds, for sizes ranging from 3 -- 20 pc.  Due to the nature of our cloud selection, our cloud sizes are limited to a finite range.  Our largest simulation is approximately 60 pc across, therefore our selected cloud regions are necessarily smaller.  There is also a lower limit to our cloud sizes as regions less than 3 pc across are deemed too small to be considered clouds for our sample (see Section~\ref{sec:sample}).   

\begin{figure} 
\begin{center}
\includegraphics[scale=0.9, angle=0, clip=true, trim=6mm 0mm 0mm 0mm]{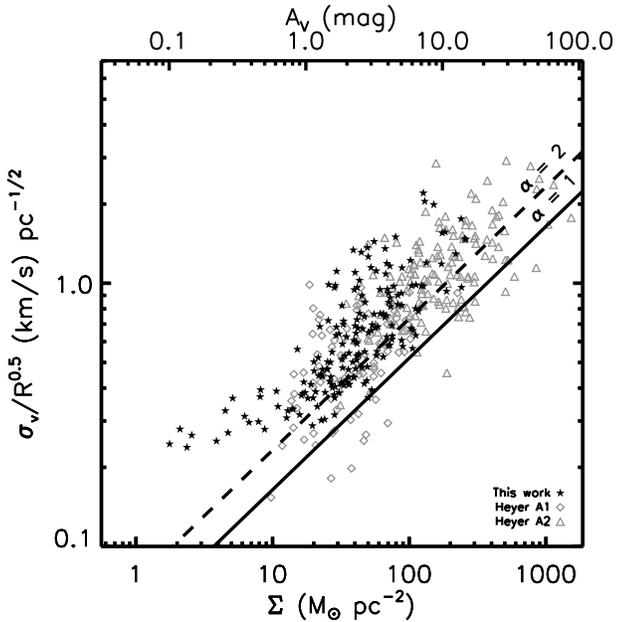}
\caption{\label{fg:heyerplot}
  Size-linewidth coefficient, $v_{\text{o}}$, as a function of cloud mass surface density, $\Sigma$, for the molecular cloud sample using \citet{solomon87} the definition of cloud size, R$_{\text{S}}$ (black stars), as compared to the observational data of \citet{heyer2009} (grey symbols).  The dashed line represents the boundary below which clouds are considered gravitationally bound ($\alpha$ = 2) and the solid line represents the boundary at which clouds are virialised ($\alpha$ = 1).  The simulated clouds follow the observed $v_{\text{o}}$-$\Sigma$ trend of \citet{heyer2009} with a Pearson correlation coefficient of 0.73.     
}
\end{center}
\end{figure}

We draw the reader's attention to the presence of large scatter about the relation in both the observational data and simulated data.  Some of this scatter can be attributed to variable cloud surface densities.  Using the size definition of Equation~\ref{eq:solrad}, we determined the size-linewidth coefficient, $v_{\text{o}} = \sigma_{1\text{D}}/\text{R$_{\text{S}}$ }^{0.5}$, for each of our clouds.  In Figure~\ref{fg:heyerplot}, we plot $v_{\text{o}}$ as a function of the cloud mass surface density, $\Sigma$ = M/$\pi$R$_{\text{S}}^2$.  We see that the surface density varies by approximately two orders of magnitude and that the scatter in this relation is greatly reduced as compared to that seen in the size-linewidth relation (Figure~\ref{fg:sizelineplot}).  Our data show an even stronger correlation between the size-linewidth coefficient and the surface density than that seen by \citet{heyer2009} with a Pearson correlation coefficient of 0.73.  Within the range of mass surface densities probed by \citet{heyer2009}, a fit to the data has the following functional form 
\begin{equation}
v_{\text{o}} \propto \Sigma^{0.49 \pm 0.04}, \label{eq:sizelwrel}
\end{equation}
as expected for clouds which are gravitationally bound and in virial equilibrium.  However, recall that this sample is populated by both bound and unbound clouds, simulated including only turbulence and gravity.  Our result indicates that the $v_{\text{o}}$-$\Sigma$ relation does not require mechanisms such as magnetic fields, variable external pressure, or stellar feedback and that the accompanying scatter may be due to the boundedness (or unboundedness) of the molecular clouds.  This interpretation of the scatter will be explored further in Section~\ref{subsec:alpha}.   

While \citet{heyer2009} measured their properties over the entirety of the cloud within an area pre-defined by \citet{solomon87}, an alternate method of determining cloud area is by counting the number of pixels in an extinction map which are above a given threshold and multiplying by the surface area of each pixel \citep[e.g.][]{lombardi2010}.  We calculated the area of our own maps in this way above the minimum column density threshold (0.5 $\times$ 10$^{21}$ cm$^{-2}$).  A primary concern for extinction-based data is the high levels of background noise in the maps.  However, since we do not have any noise in our simulations, we can analyse our synthetic observations as extinction maps without concern of contamination from the background.  The mass surface density is therefore equal to the mass above the threshold, M, divided by the surface area above the threshold, A, and the effective radius, R$_{\text{L}}$, is 
\begin{equation}
	\text{R}_{\text{L}} = \sqrt{\frac{\text{A}}{\pi}}.
\label{eq:rlom}
\end{equation}   

\begin{figure} 
\begin{center}
\includegraphics[scale=0.9, angle=0, clip=true, trim=8mm 0mm 0mm 0mm]{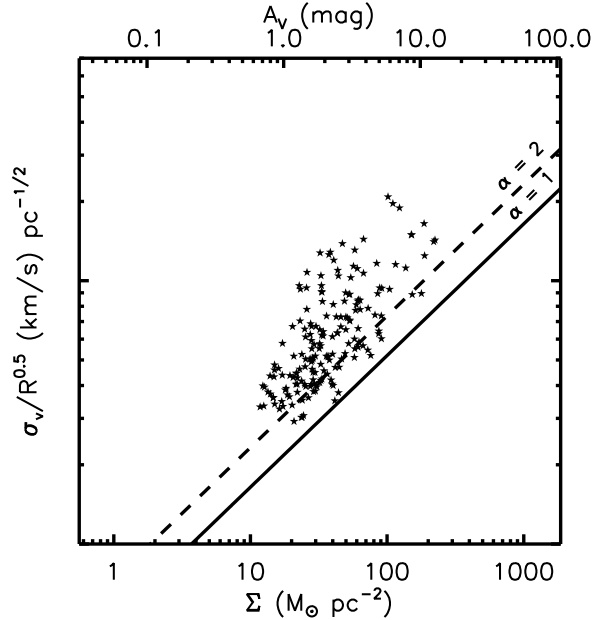}
\caption{\label{fg:lombardiplot}
  Size-linewidth coefficient, $v_{\text{o}}$, as a function of cloud mass surface density, $\Sigma$, for the molecular cloud sample using the \citet{lombardi2010} definition of cloud size, R$_{\text{L}}$.  The boundaries below which clouds are considered gravitationally bound (dashed line, $\alpha$ = 2) and virialised (solid line, $\alpha$ = 1) are also shown.
}
\end{center}
\end{figure}

Using this definition of cloud size, we re-derived our cloud properties.  Figure~\ref{fg:lombardiplot} shows $v_{\text{o}} = \sigma_{1\text{D}}/\text{R$_{\text{L}}$ }^{0.5}$, as a function of $\Sigma$ = M/$\pi$R$_{\text{L}}^2$.  A fit to the data plotted in Figure~\ref{fg:lombardiplot} shows that the size-linewidth coefficient, $v_{\text{o}}$, is again proportional to $\Sigma^{1/2}$ as expected from Equation~\ref{eq:coefficient}.  We also calculated the mean surface density to be $\bar{\Sigma}$ = 38 $\pm$ 4 M$_{\odot}$ pc$^{-2}$, which is consistent with the median surface density of 42 M$_{\odot}$ pc$^{-2}$ reported by \citet{heyer2009}.  

\citet{lombardi2010} found from their observations that cloud column densities are \emph{constant} above a given threshold and that their value is dependent on the threshold.  We tested this result for our clouds by recalculating the cloud properties for three higher thresholds (A$_{\text{v}}$ $\ge$ 1).  We find that if a cloud is probed to very low extinctions, the $v_{\text{o}}$-$\Sigma$ relation can be recovered; otherwise, the surface density will appear to be roughly constant since the range of surface densities becomes limited \citep[e.g.][]{lombardi2010,kauf2010a}.  Since the lower limit for detection in current observations is A$_{\text{v}}$ $\sim$ 0.5 \citep{beaumont2012, kain2009}, extinction-based data will naturally result in a constant column density since the threshold is necessarily too high (typically A$_{\text{v}}$ $>$ 1).

\subsection{Determining an appropriate size estimator}\label{subsec:size}

We have explored two methods for defining cloud size which give similar results.  However, which is the best way to characterise a cloud's size?  A definition of size that provides an accurate estimate of the gravitational potential energy is particularly useful in order to estimate the virial parameter.

Using three-dimensional information from the simulations we can define a cloud's size using the gravitational potential \citep{smith2009, gong2011}.  We can determine the gravitational potential energy, U, for each cloud in our sample as follows,
\begin{equation}
	\text{U} = \sum_i{\frac{1}{2}\text{m$_i$}\phi_i},
\label{eq:pot}
\end{equation}
where m$_{\text{i}}$ is the mass and $\phi_i$ is the gravitational potential of each particle in the cloud.  As mentioned previously, we only included particles contained in pixels above a minimum column density threshold of 0.5$\times$10$^{21}$ cm$^{-2}$ in our determination of cloud properties.  We have also included sink particles in our calculation of the gravitational potential energy and total mass of the cloud.  Using the gravitational potential energy, we then derive a radius, R$_{\text{pot}}$:
\begin{equation}
	\text{R}_{\text{pot}} = \frac{3}{5}\frac{\text{GM}^{2}}{\left|\text{U}\right|},
\end{equation}
where M is the total mass of the cloud and stars.  We find that our estimate of the radius, R$_{\text{pot}}$, is not affected by the inclusion of sink particles.  The factor of 3/5 assumes that the clouds can be represented as homogeneous spheres, which is a common assumption when deriving properties of molecular clouds.  This coefficient varies by less than 25 percent for clouds that are more accurately represented as 1:2 prolate or oblate homogeneous ellipsoids and we conclude that a sphere is a fair approximation for the global shape of a molecular cloud.  

\begin{figure} 
\begin{center}
\includegraphics[scale=0.85, angle=0, clip=true, trim=1.3cm 0cm 0cm 0cm]{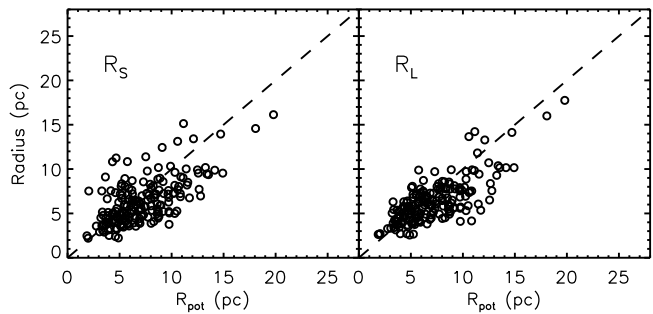}
\caption{\label{fg:radiuscomp}
  Observed radius as a function of R$_{\text{pot}}$.  The dashed one-to-one line is also shown.  The observed radii are the effective radius from \citet{solomon87} (left) and the effective radius from \citet{lombardi2010} (right).
  }
\end{center}
\end{figure}

If we compare R$_{\text{pot}}$ to the estimates for cloud radius derived from the two methods that we have previously explored, we can determine the best observational estimate of cloud size for the specified purpose of finding the boundedness of a cloud.  Figure~\ref{fg:radiuscomp} shows the two observational estimates of radius plotted as a function of R$_{\text{pot}}$.  Both observed radii are quite representative of the full radius of the cloud as determined by the gravitational potential energy.  The radii as determined by the method of \citet{lombardi2010} are slightly more correlated with R$_{\text{pot}}$ with a Pearson coefficient of 0.75 compared to a Pearson coefficient of 0.65 for the radii as determined by the \citet{solomon87} method.  The principal variation from cloud to cloud is its shape which gives rise to the scatter seen in Figure~\ref{fg:radiuscomp}.  Clouds which are highly non-spherical will result in an overestimation of cloud size by the method of \citet{solomon87}.  These results indicate that the method for determining the cloud radius outlined in \citet{lombardi2010} is a more useful means of characterising the size of a cloud with respect to its virial state.

\subsection{Do unbound clouds match observations?}\label{subsec:alpha}

In the previous section, we determined that the most appropriate measure of a cloud's size can be made using the method of \citet{lombardi2010}.  From the masses, linewidths, and sizes as defined by \citet{lombardi2010} for the molecular clouds in our sample, we determine the virial parameter of each cloud.  We note that a large fraction of the clouds are strictly unbound ($\alpha$ $>$ 2).  

In Section~\ref{subsec:cloudprops}, we interpreted the scatter in Figures~\ref{fg:heyerplot} and \ref{fg:lombardiplot} as being due to a spread in the virial parameter.  By definition, the size-linewidth coefficient not only depends on the mass surface density, but it is also dependent on the virial parameter, $\alpha$: 
\begin{equation}
		v_{\text{o}} = (\pi \text{G} \alpha \Sigma/5)^{1/2}. \label{eq:newsizelinecoeff}
\end{equation}
However, the virial parameter is commonly assumed to be a constant rather than a property capable of variation.  \citet{heyer2009} assumed that $\alpha$ = 1 and thus disregarded the virial parameter from their arguments concerning the scatter in the relation between the size-linewidth coefficient and the surface density.  If we consider the idea that clouds may not necessarily be gravitationally bound or virialised ($\alpha$ $\neq$ 1, see Section~\ref{sec:Intro}), the scatter can be explained by the boundedness (or unboundedness) of the molecular cloud.  

In order to determine whether the observed virial parameter defined by Equation~\ref{eq:alphadef} is actually reflective of the true virial state of the cloud, we calculated the physical virial parameter given by
\begin{equation}
	\alpha_{\text{3D}} = 2K/U,
\end{equation}
where U is the gravitational potential energy determined by Equation~\ref{eq:pot} and K is the kinetic energy determined from the mass of the cloud and stars, $M_{\text{tot}}$, and one-dimensional velocity dispersion, $\sigma_{\text{1D}}$, such that
\begin{equation}
	K = \frac{3}{2}M_{\text{tot}}\sigma_{\text{1D}}^2.
\end{equation}

\begin{figure} 
\begin{center}
\includegraphics[scale=1, angle=0]{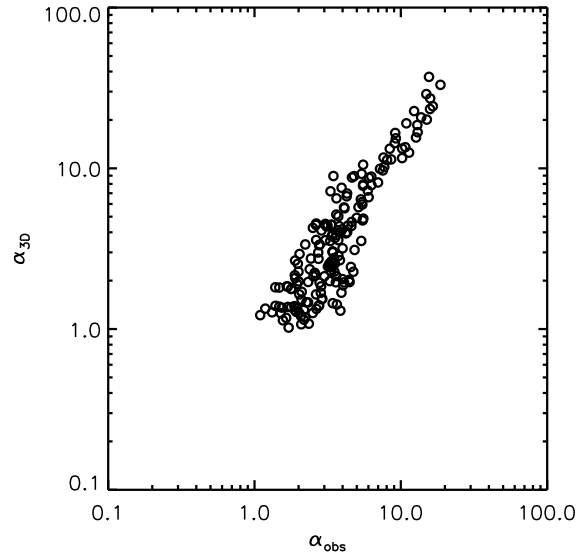}
\caption{\label{fg:alphacomp}
True physical virial parameter, $\alpha_{\text{3D}}$ = 2K/U, as a function of the observed virial parameter, $\alpha_{\text{obs}}$ = 5$\sigma^2$R/GM. }
\end{center}
\end{figure}

Figure~\ref{fg:alphacomp} shows the tight correlation between the observed and physical virial parameters, indicating that many of the molecular clouds in our sample ($>$ 69\%) are truly unbound with $\alpha_{\text{3D}}$ $>$ 2.  Therefore, a measure of the virial parameter using an appropriate size estimator is representative of the true virial state of the molecular cloud, which characterises the amount of scatter in the $v_{\text{o}}$-$\Sigma$ relation.  

In previous sections, we found that our clouds, which range from bound to unbound as defined by their true virial state, match the observational trends.  As shown above, we also find that observational estimates of cloud virial parameters should be close to the true values.   It is thus reasonable to take the observations at face value and assume that a substantial fraction of real molecular clouds are unbound.

\section{Conclusions \& Discussion}\label{sec:discussion}

We studied the properties of a sample of molecular clouds with masses ranging from $\sim$ 300 -- 10 000 M$_{\odot}$,  which were formed within larger volumes and simulated with only turbulence and gravity.  This sample follows the observed trend of the size-linewidth coefficient, $v_{\text{o}}$ = $\sigma_{v}$/R$^{1/2}$, with the surface density, $\Sigma$, from \citet{heyer2009}.   We also argued that the virial state, and consequently the boundedness, of a molecular cloud is the source of the scatter which accompanies the trend.  We have shown that the observed properties can be understood if molecular clouds are a population of turbulent, transient, and often unbound structures.  This interpretation is consistent with the data of both \citet{heyer2009} and \citet{larson81}.  These findings support the idea that clouds need not be maintained in dynamical equilibrium ($\alpha=1$).  Recent work by \citet{kritsuk13}, using periodic mesh simulations, reached similar conclusions.  This implies that additional support from magnetic fields, variable external pressure, or stellar feedback may be less important than is commonly assumed.

Assuming virialized clouds is likely to provide erroneous mass estimates.  However, if we cannot assume virialization, this raises the question of how to estimate $\alpha$ for a molecular cloud based on observable quantities.   We noted in Section~\ref{subsec:cloudprops} that the surface density is relatively constant above high thresholds, consistent with the results of \citet{lombardi2010} and \citet{kauf2010a}.  Therefore, provided that we can measure the size and linewidth of a molecular cloud, we can solve for the virial parameter, $\alpha$, and its mass, assuming a functional dependence of the column density on the chosen threshold \citep[see also][]{balle2012}.  We find that the estimate of cloud size as determined from extinction maps \citep[e.g.][]{lombardi2010} provides an accurate way to determine the cloud's gravitational potential energy and, subsequently, its virial parameter.

We find that all clouds in our sample ultimately form collapsing cores, regardless of whether the cloud is bound, consistent with the results of \citet{clark05}.  Localised regions of compression due to turbulence within unbound clouds lead to isolated regions of star formation as the clouds disperse and their column density fades beneath the lower limit for detection.  We conclude that molecular clouds do not have to be bound to form stars or to have observed properties like those of nearby low-mass clouds.  If molecular clouds are unbound, this allows for relatively low star formation rates for individual clouds, matching those seen in observations.  These low star formation rates are naturally produced for unbound clouds as they are not directly tied to their free-fall times.  We will explore the evolution of cloud properties, including the star formation rate, in future work.   Since our simulations are roughly scale-free, we expect our results to apply to clouds of higher masses as well.  A substantial role for unbound clouds could help resolve the long standing issue of low galactic star formation rates despite the abundance of molecular gas.

\section*{Acknowledgments}

We would like to thank SHARCNET (Shared Hierarchical Academic Research Computing Network) 
and Compute/Calcul Canada, which provided dedicated resources to run these simulations.   
This work was supported by NSERC.  J.W. acknowledges support from the Ontario Early Researcher Award (ERA).
The authors would like to thank the anonymous referee for useful comments that have improved this work.

\bibliographystyle{apj}

\end{document}